\documentclass[aps,english,twocolumn,superscriptaddress]{revtex4-1}

\usepackage{longtable}
\usepackage{morefloats}
\usepackage[dvips]{graphicx}
\usepackage{color}
\usepackage{epsfig,graphicx,amsfonts,amsbsy}
\usepackage{amsmath,amsfonts,amsthm,amssymb}
\usepackage{appendix}
\usepackage{makeidx}
\usepackage{url}
\usepackage{verbatim}
\usepackage[bookmarksnumbered,pdfpagelabels=true,plainpages=false,colorlinks=true,linkcolor=blue,citecolor=red,urlcolor=blue]{hyperref}
\usepackage[rightcaption]{sidecap}
\usepackage{array}
\usepackage{booktabs}
\usepackage{multirow}
\usepackage{tabularx}
\usepackage{cancel,soul,ulem}

\newcommand{\bs}[1]{\boldsymbol{#1}}
\newcommand{\ssf}[1]{\mathsf{#1}}

\makeatletter

\begin{document}

\title{\bf{Controlling skyrmion helicity via engineered Dzyaloshinskii-Moriya interactions}}
\author{Sebasti\'an A. D\'iaz}
\affiliation{Department of Physics, University of California, San Diego, La Jolla, California 92093, USA}
\author{Roberto E. Troncoso}
\affiliation{Departamento de F\'isica, Universidad T\'ecnica Federico Santa Mar\'ia, Avenida Espa\~na 1680, Valpara\'iso, Chile}

\begin{abstract}
Single magnetic skyrmion dynamics in chiral magnets with a spatially inhomogeneous Dzyaloshinskii-Moriya interaction (DMI) is considered. Based on the relation between DMI coupling and skyrmion helicity, it is argued that the latter must be included as an extra degree of freedom in the dynamics of skyrmions. An effective description of the skyrmion dynamics for an arbitrary inhomogeneous DMI coupling is obtained through the collective coordinates method. The resulting generalized Thiele's equation is a dynamical system for the center of mass position and helicity of the skyrmion. It is found that the dissipative tensor and hence the Hall angle become helicity dependent. The skyrmion position and helicity dynamics are fully characterized by our model in two particular examples of engineered DMI coupling: half-planes with opposite-sign DMI and linearly varying DMI. In light of the experiment of Shibata {\it et al}. [Nature Nanotech. {\bf 8}, 723 (2013)] on the magnitude and sign of the DMI, our results constitute the first step toward a more complete understanding of the skyrmion helicity as a new degree of freedom that could be harnessed in future high-density magnetic storage and logic devices. 
\end{abstract}


\maketitle
\section{Introduction} 

Skyrmions, swirling spin textures appearing in chiral magnets, are magnetic structures in which the spins point in all the directions wrapping a sphere\cite{Skyrme,Bogdanov}. These topological spin textures have ignited a growing interest in spintronics\cite{nagaosa} due to their rich phenomenology as well as novel potential applications. Their nano-scale size, topologically-protected stability\cite{Bogdanov}, and the very low electric current densities needed to displace them \cite{Jonietz} are among the best qualities that make them attractive candidates for information carriers in high-density data-storage technologies\cite{Fert1}. They have been explored in bulk magnets like MnSi\cite{Muhlbauer,Jonietz,Neubauer}, in thin films of Fe$_{1-x}$Co$_x$Si\cite{Munzer,PMilde,Pfleiderer,Yu2}, Mn$_{1-x}$Fe$_x$Ge\cite{Shibata}, FeGe\cite{Yu1}, La$_{0.5}$Ba$_{0.5}$MnO$_{3}$\cite{Nagao}, CuOSeO$_{3}$\cite{Seki}, and also in a Fe monoatomic layer on Ir(111)\cite{Heinze}. Controlling the temperature and the external magnetic field applied to the samples, spontaneous skyrmion phases have been realized and detected by neutron scattering\cite{Muhlbauer}, Lorentz transmission electron microscopy (LTEM)\cite{Yu2}, and spin-resolved scanning tunneling microscopy\cite{Heinze} experiments.

Current-driven skyrmion dynamics also displays intriguing topological transport properties\cite{Neubauer,Shiomi,Everschor}, phenomena that result from the so-called spin-transfer torques exerted by carrier spins on the magnetization\cite{Berger}. Other mechanisms such as thermal gradients\cite{Mochizuki}, inhomogeneity in the fields\cite{Fert,Troncoso,Reichhardt}, and magnon currents\cite{Lin1} have also been proposed to induce the motion of skyrmions. In these studies it was assumed that the evolution of the skyrmion magnetization could be appropriately treated using a reduced set of collective coordinates. Under this approach the Landau-Lifshitz-Gilbert (LLG) equation is mapped to a particle-like equation of motion, also known as Thiele's equation\cite{Thiele}.

The spin texture of magnetic skyrmions originates from the Dzyaloshinskii-Moriya interaction (DMI)\cite{DMI}. This interaction not only plays a fundamental role in the nucleation and stability of skyrmions, but also in the helical and conical phases\cite{Bogdanov}. For instance, properties like the size and helicity of skyrmions are set by the magnitude and sign of the DMI coupling, respectively. It has been shown that single skyrmions can be nucleated by injection of a spin-polarized current\cite{Heinze,Fert} or local heating \cite{Koshibae} on magnetic thin-films. Magnetic skyrmions may emerge as stable vortex- or hedgehog-type spin configurations\cite{Bogdanov} depending on the type of DMI, which could be due to bulk interactions or induced at the interface of magnetic films in contact with heavy metals\cite{Fert}. Vortex-type (or Bloch-type) skyrmions, those for which the magnetization swirls around their center and perpendicular to their radial direction, have been observed in non-centrosymmetric magnets with B20-type crystal structure such as Mn$_{1-x}$Fe$_x$Ge\cite{Shibata}. Meanwhile, hedgehog-type (or N\'eel-type) skyrmions, whose spins point either radially outward or inward, have been observed in magnetic ultra-thin films of Fe/Ir(111)\cite{Heinze} or in the recently reported polar magnetic semiconductor GaV$_4$S$_8$\cite{Tsurkan}.

Recent experiments have revealed the dependence of the DMI in the chiral magnet alloys Mn$_{1-x}$Fe$_{x}$Ge\cite{Shibata,Grigoriev} and Fe$_{1-x}$Co$_x$Si \cite{Grigoriev2,Siegfried,Morikawa} on the chemical composition $x$. For instance, measuring the wave vector of the helical spin texture, it has been shown\cite{Siegfried} that the strength and even the sign of the DMI can be controlled by varying the composition of Co in Fe$_{1-x}$Co$_x$Si. Similar results were reported in Ref. [\onlinecite{Shibata}], where by changing the composition $x$ in Mn$_{1-x}$Fe$_{x}$Ge the size and helicity of magnetic skyrmions were modified. These experiments, together with the latest theoretical studies\cite{Koretsune,Gayles,Chen}, have been devoted to elucidate the intriguing relation of the magnitude and sign of the DMI to chemical composition, thus  
opening the possibility to manipulate the skyrmion helicity by engineering the crystal chirality and spin-orbit coupling. Motivated by these experiments\cite{Shibata,Grigoriev,Grigoriev2,Siegfried,Morikawa}, here we study the dependence of the helicity degree of freedom on single magnetic skyrmion dynamics in chiral magnets with an inhomogeneous DMI coupling.

The structure of this paper is the following. In section II, an effective description of the skyrmion dynamics based on the collective coordinates method is developed assuming a general inhomogeneous DMI coupling. The generalized Thiele's equation thus obtained, an autonomous dynamical system for the skyrmion center of mass position and helicity, is then further reduced to the case of a DMI coupling with arbitrary one-dimensional spatial dependence. Two particular cases are then analyzed taking advantage of the general results derived in this section. Sections III and IV dwell upon half-planes with opposite-sign DMI coupling and linearly varying DMI coupling, respectively. The role of the helicity and the set of possible trajectories allowed by the dynamical systems describing each case are determined. Finally, in section V, these two cases are compared, possible measurements that could be performed to confirm our predictions are briefly discussed, and our conclusions are presented. 
 
\section{Effective Skyrmion Dynamics}

The time evolution of any magnetic texture, such as a skyrmion, can be determined by solving the field equations for the magnetization field $\bs{M}$. However, in the systems we address in this paper, not all of the uncountably infinite degrees of freedom in the magnetization field are relevant. For a static, axially-symmetric skyrmion centered at the origin, the direction of the magnetization, $\bs{\Omega} \equiv \bs{M}/M$, can be written as
\begin{eqnarray}
\nonumber\bs{\Omega}(\bs{r}, \gamma) &=& \sin\Theta(r, \gamma) [ \cos(\varphi + \gamma)\hat{\bs{x}} + \sin(\varphi + \gamma)\hat{\bs{y}} ]\\
\label{eq:StaticSk}&& + \cos\Theta(r,\gamma)\hat{\bs{z}},
\end{eqnarray}
where $r$ and $\varphi$ are the usual cylindrical coordinates. It then follows that to completely specify this magnetic texture, only its helicity $\gamma$ and radial profile $\Theta(r, \gamma)$ are required. At low temperatures and in the presence of an external magnetic field $\bs{B}$, these defining quantities can be obtained by minimizing the corresponding magnetic energy functional $\mathcal{U}$ which, up to an overall constant, is given by
\begin{equation}\label{eq:MagneticEnergy}
\mathcal{U} = M^2\!\!\int\!\! d\bs{r}\! \left[ J(\nabla\bs{\Omega})^2 + 2D(\bs{r})\bs{\Omega}\cdot\nabla\times\bs{\Omega}  - \bar{\bs{B}}\cdot\bs{\Omega} \right],
\end{equation}
where $J$ denotes the spin stiffness and $\bar{\bs{B}} = \bs{B}/M$. In all the results presented here $\bs{B} = B\bs{\hat{z}}$, i.e., orthogonal to the sample. This model assumes the magnitude of the magnetization field is uniform across the sample, it neglects anisotropy energies, and allows for a spatially-dependent DMI coupling $D(\bs{r})$. It has been successfully used to describe skyrmions in chiral magnet thin films with uniform DMI coupling, i.e., when $D(\bs{r}) = D$\cite{Yu2}.

The physical scenario we wish to study begins with a skyrmion already nucleated in a region of the sample with a locally uniform DMI coupling. Although this magnetic texture may get distorted as it evolves, owing to its topologically-protected stability, it will remain being a skyrmion. Furthermore, for low energy distortions, which correspond to long-term dynamics, the axial symmetry is retained\cite{JRoldan}. Therefore, we expect the skyrmion to translate across the sample and, since the helicity is inextricably related to the DMI, its helicity is also expected to evolve in time as the skyrmion explores regions with varying DMI. The previous analysis and assumptions justify an effective description of the skyrmion dynamics in terms of the position of its center $\bs{r}_0 = X\bs{\hat{x}} + Y\bs{\hat{y}}$ and its helicity $\gamma$.

\subsection{Collective Coordinates and Generalized Thiele's Equation}\label{sec:Coordinates}

Now that we have identified the relevant degrees of freedom for the effective long-term skyrmion dynamics, we need to determine their equations of motion. To that end we will use the method of collective coordinates\cite{Rajaraman} and closely following Ref. \cite{Tretiakov1} we will construct the corresponding Generalized Thiele's Equation. 

The method of collective coordinates is based on the assumption that the magnetic texture can be described by a discrete set of time-dependent coordinates $\bs{\xi}(t)^T = (\xi_1(t),\dots,\xi_N(t))$, so that the time dependence of the magnetization comes solely from them, i.e., $\bs{M}(\bs{r}, t)=\bs{M}( \bs{r}, \bs{\xi}(t))$. For our effective skyrmion dynamics, as discussed above, we have three collective coordinates: $\xi_1(t) = X(t)$, $\xi_2(t) = Y(t)$, and $\xi_3(t) = \gamma(t)$. When the magnitude of the magnetization can be regarded as constant, the equations of motion for the collective coordinates follow from the dynamics of the direction of the magnetization $\bs{\Omega}$, governed by the Landau-Lifshitz-Gilbert (LLG) equation
\begin{equation}
\frac{d \bf{\Omega}}{dt} = \bar{\gamma} {\bf H}_{\scriptsize\mbox{eff}}\times{\bf \Omega} + \alpha{\bf \Omega}\times\frac{d {\bf \Omega}}{dt}+{\bs \tau}_{\text{STT}}+{\bs \tau}_{\text{SOT}}.
\end{equation}
Here $\bar{\gamma} = g\frac{|e|}{2m_e}$ is the electron gyromagnetic ratio, the effective magnetic field is ${\bf H}_{\scriptsize\mbox{eff}}= - \frac{1}{M}\frac{\delta \mathcal{U}}{\delta{\bf \Omega}}$, and $\alpha$ is the Gilbert damping constant. For completeness, the conventional spin-transfer torque ${\bs \tau}_{\text{STT}}$, due to the transference of spin angular momentum from spin-polarized currents to the magnetization, and a spin-orbit torque ${\bs \tau}_{\text{SOT}}$\cite{Hals}, due to the intrinsic spin-orbit coupling of the chiral magnet, have also been included. Using that the time dependence of $\bs{\Omega}$ enters through the collective coordinates, i.e., $\bs{\Omega}(\bs{r}, t)=\bs{\Omega}(\bs{r}, \bs{\xi}(t))$, the left-hand side of this equation can be rewritten as $d\bs{\Omega}/dt = \dot{\xi_j}\partial\bs{\Omega}/\partial\xi_j$. Acting on the LLG equation with $\bs{\Omega}\times$, taking the dot product with $\partial\bs{\Omega}/\partial\xi_i$, multiplying by the angular momentum density $\mathsf{J}=M/\bar{\gamma}$, and finally integrating over the volume, we arrive at the Generalized Thiele's Equation 
\begin{equation}\label{eq:ThieleEquation}
G_{ij}{\dot\xi}_j - \mathcal{D}_{ij}\dot{\xi}_j + F_i + F^s_i = 0,
\end{equation}

\begin{figure*}[t]
\includegraphics[width=0.98\textwidth]{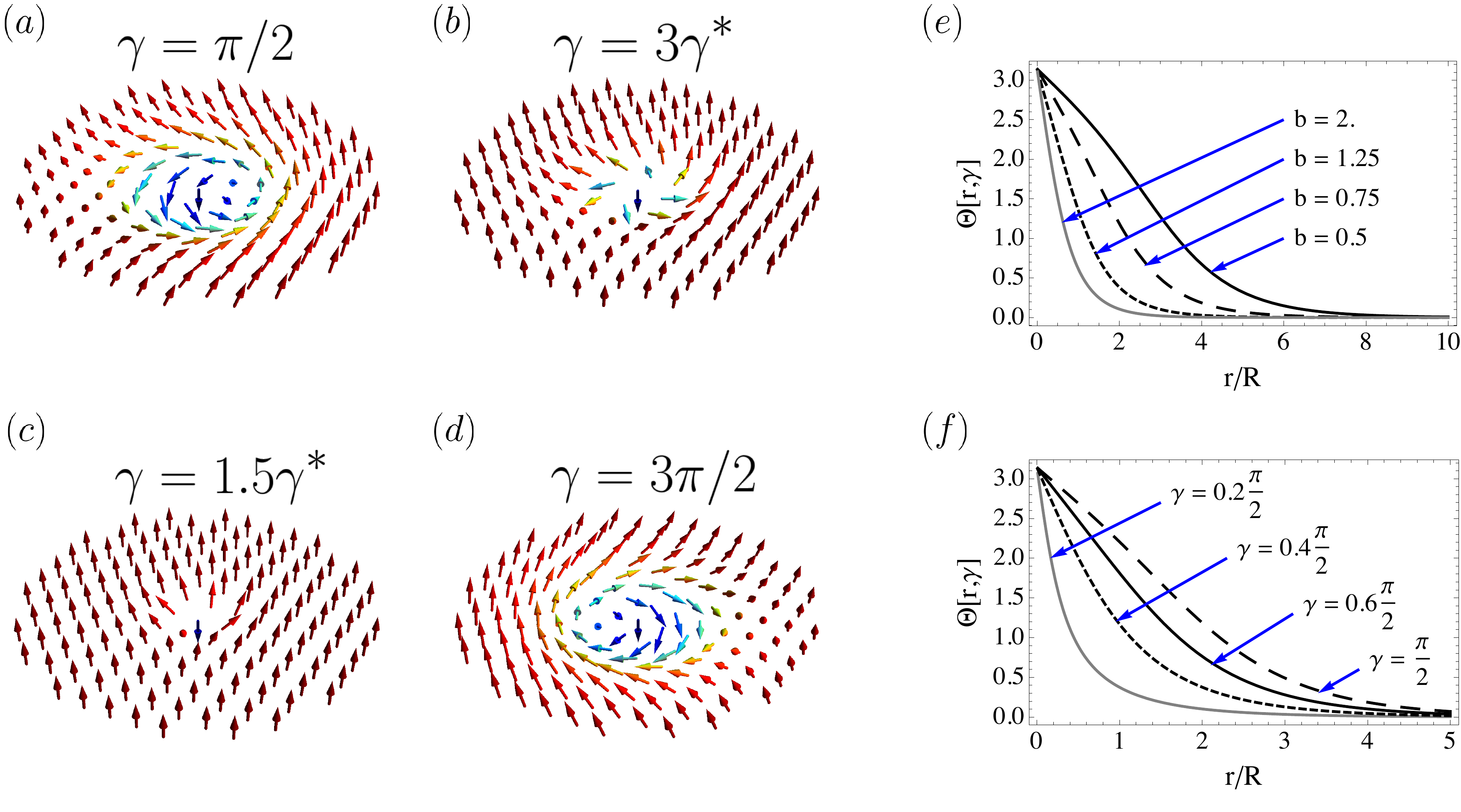}
\caption{(color online). Schematic illustration of skyrmion textures with different helicities are displayed in (a)-(d), for $\gamma=\pi/2$, $\gamma=3\gamma^*$, $\gamma=1.5\gamma^*$ and $\gamma=3\pi/2$, respectively (here $b = 2$ and $\rho^* = 0.04$ which yield $\gamma^* = 0.23$). In (e) and (f), plots of the numerical solutions of the adimensionalized profile equation, Eq. (\ref{eq:skyrmionprofile}), for a single skyrmion are shown. These helicity-dependent solutions were determined in the presence of an adimensionalized uniform external magnetic field $b$. In (e), where the helicity of the skyrmion profile is $\gamma=\pi/2$, it is observed that as the magnetic field increases, the size of the skyrmion core shrinks. A similar behavior holds for the skyrmion size as the helicity decreases, which is plotted in (f) for a magnetic field $b=0.75$. The size of the skyrmion core is a smooth function of the helicity, being maximum when $\gamma = \pi/2$ (a) or $\gamma = 3\pi/2$ (d) and minimum for $\gamma \in [0,\gamma^*]\cup[\pi - \gamma^*,\pi + \gamma^*]\cup[2\pi - \gamma^*,2\pi]$.}
\label{fig:skyrmiontextures}
\end{figure*}
where
\begin{align}
G_{ij} &\label{eq:skyrmionforces1} = \ssf{J}\int d\bs{r}\;\bs{\Omega}\cdot \left( \frac{\partial\bs{\Omega}}{\partial\xi_i} \times \frac{\partial\bs{\Omega}}{\partial\xi_j} \right),\\
{\cal D}_{ij} &\label{eq:skyrmionforces2} = \alpha\ssf{J}\int d\bs{r} \left( \frac{\partial\bs{\Omega}}{\partial\xi_i} \cdot \frac{\partial\bs{\Omega}}{\partial\xi_j} \right),\\
F_i &\label{eq:skyrmionforces3} = - \int d\bs{r}\frac{\delta\mathcal{U}}{\delta\bs{\Omega}}\cdot\frac{\partial\bs{\Omega}}{\partial\xi_i} = - \frac{\partial\mathcal{U}}{\partial\xi_i},\\
F^s_i &\label{eq:skyrmionforces4}= -\mathsf{J}\int d\bs{r}\frac{\partial\bs{\Omega} }{\partial\xi_i}\cdot\left[\bs{\Omega}\times\left({\bs \tau}_{\text{STT}}+{\bs \tau}_{\text{SOT}}\right)\right].
\end{align}
Above, $G_{ij}$ and $\mathcal{D}_{ij}$ denote the gyrotropic and dissipative tensor, respectively. $\bs{F}$ is called the generalized force. The spin force, $\bs{F}^s$, is determined by the two-dimensional electric current density and the specific form of ${\bs \tau}_{\text{SOT}}$. In the results presented here the electric current density is assumed to vanish, so this force will be neglected. Its role in skyrmion dynamics in systems with spatially inhomogeneous DMI will be discussed elsewhere\cite{work in preparation}. By solving the Generalized Thiele's Equation, we obtain the time dependence of the collective coordinates necessary to describe the effective skyrmion dynamics. In order to do so, we first need to compute $G_{ij}$, $\mathcal{D}_{ij}$, and $\bs{F}$ which we proceed to do below.

\subsection{Gyrotropic and Dissipative Tensors}

As can be seen from their definition, while the gyrotropic tensor is antisymmetric, the dissipative tensor is symmetric. Moreover, these two tensors depend only on the collective coordinate-dependent magnetic texture. The magnetic energy $\mathcal{U}$ plays absolutely no role in their calculation. From our previous discussion, we will use the magnetic texture introduced above Eq. \eqref{eq:StaticSk}, but with its center and helicity now promoted to collective coordinate status: $\bs{\Omega}(\bs{r}, t) = \bs{\Omega}(\bs{r} - \bs{r}_0(t), \gamma(t))$. 

We found that the non-zero components of the gyrotropic tensor are $G_{12} = - G_{21} = 4\pi d\ssf{J}$, with $d$ the thickness of the magnetic film. Noting that $G_{ij}$ is proportional to the topological charge in the $\xi_i\xi_j$ space, this result follows directly from the topological structure inherited from the parent magnetic texture $\bs{\Omega}(\bs{r}, \gamma)$. Our calculations show that the dissipative tensor obeys $\mathcal{D}_{ij} = 4\pi \alpha d\ssf{J}\delta_{ij}\eta_j$, with $\eta_1=\eta_2$ by the axial symmetry of the parent texture, and where 
\begin{align}
\eta_1(\gamma)&\label{eq:eta1}=\frac{1}{4}\int^{\infty}_0 rdr\left[\left(\frac{\partial\Theta(r,\gamma)}{\partial r}\right)^2+\frac{\sin^2\Theta(r,\gamma)}{r^2}\right],\\
\eta_3(\gamma)&\label{eq:eta3}=\frac{1}{2}\int^{\infty}_0 rdr\left[\left(\frac{\partial\Theta(r,\gamma)}{\partial\gamma}\right)^2+\sin^2\Theta(r,\gamma)\right].
\end{align}
Clearly $\eta_1, \eta_2, \eta_3 > 0$, so $\mathcal{D}_{ij}$ is positive definite. The radial profile $\Theta(r,\gamma)$, required to compute the $\eta_i$'s, was obtained by numerically solving the corresponding Euler-Lagrange equation. Details of the calculation of the skyrmion radial profile are outlined in Appendix \ref{appendix}. 

In Fig. \ref{fig:skyrmiontextures} the skyrmion textures and profiles for different helicities and external magnetic fields are displayed. The magnetization texture is illustrated in panels (a)-(d), for $\gamma=\pi/2$, $\gamma=3\gamma^*$, $\gamma=1.5\gamma^*$ and $\gamma=3\pi/2$, respectively.  A minimum value for the helicity, which is denoted by $\gamma^*$, has been introduced in our continuum model to account for the discrete nature of the magnetic skyrmion. Because the size of the skyrmion can not be less than the lattice constant, an elementary size for the skyrmion is considered and accordingly, a minimum value for the helicity is determined. This value is estimated for Fe$_{0.5}$Co$_{0.5}$Si \cite{Grigoriev2} to be $\gamma^* = 0.23$. Meanwhile, in (e) and (f), plots of helicity-dependent solutions of the adimensionalized profile equation, Eq. (\ref{eq:skyrmionprofile}), for a single skyrmion in the presence of an adimensionalized uniform external magnetic field $b=J\bar{B}/2D^2$ are shown. In (e), for a helicity $\gamma=\pi/2$, it is observed that as the magnetic field increases, the size of the skyrmion core shrinks. A similar behavior holds for the skyrmion size, plotted in (f) for a magnetic field $b=0.75$, as the helicity decreases. The size of the skyrmion core is a smooth function of the helicity, being maximum when $\gamma = \pi/2$ (a) or $\gamma = 3\pi/2$ (d) and minimum for $\gamma \in [0,\gamma^*]\cup[\pi - \gamma^*,\pi + \gamma^*]\cup[2\pi - \gamma^*,2\pi]$.

\begin{figure}[ht]
\begin{center}
\includegraphics[width=1.62in,height=1.08in]{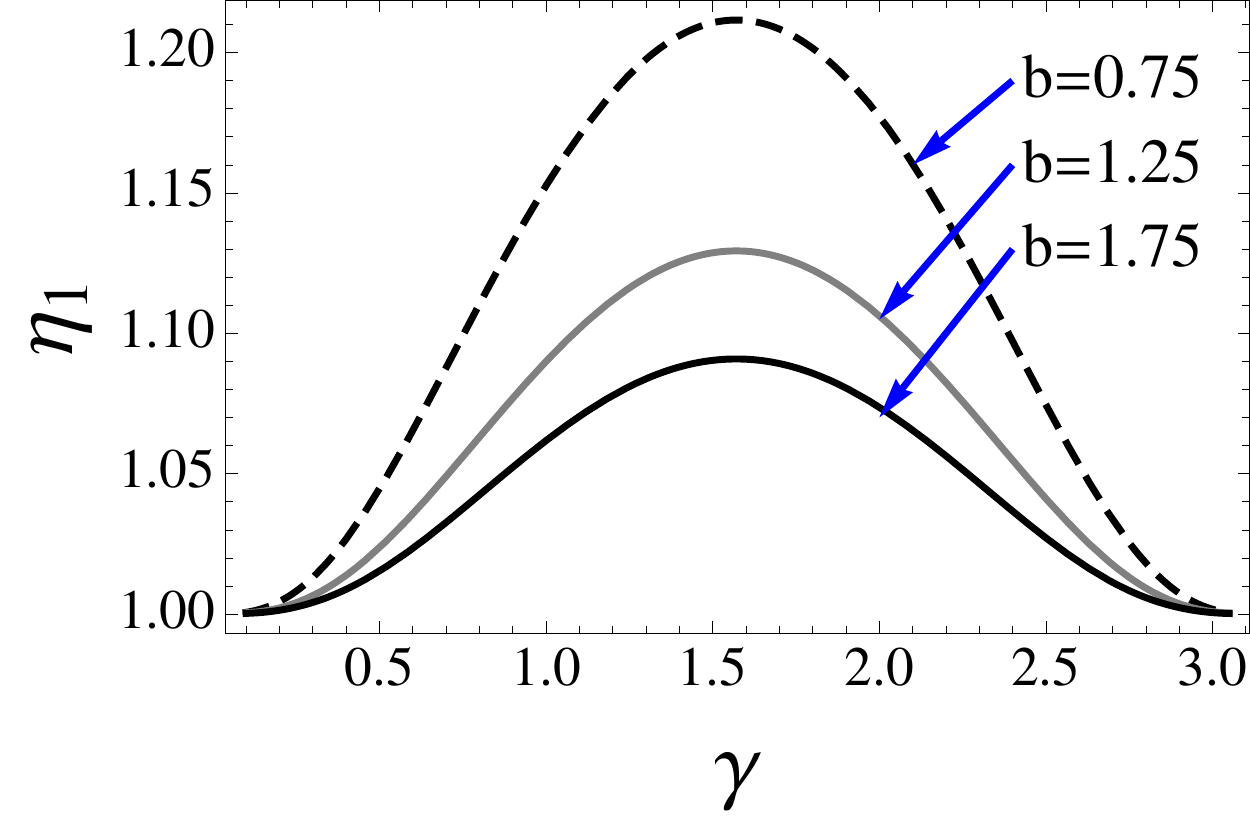}
\includegraphics[width=1.64in,height=1.08in]{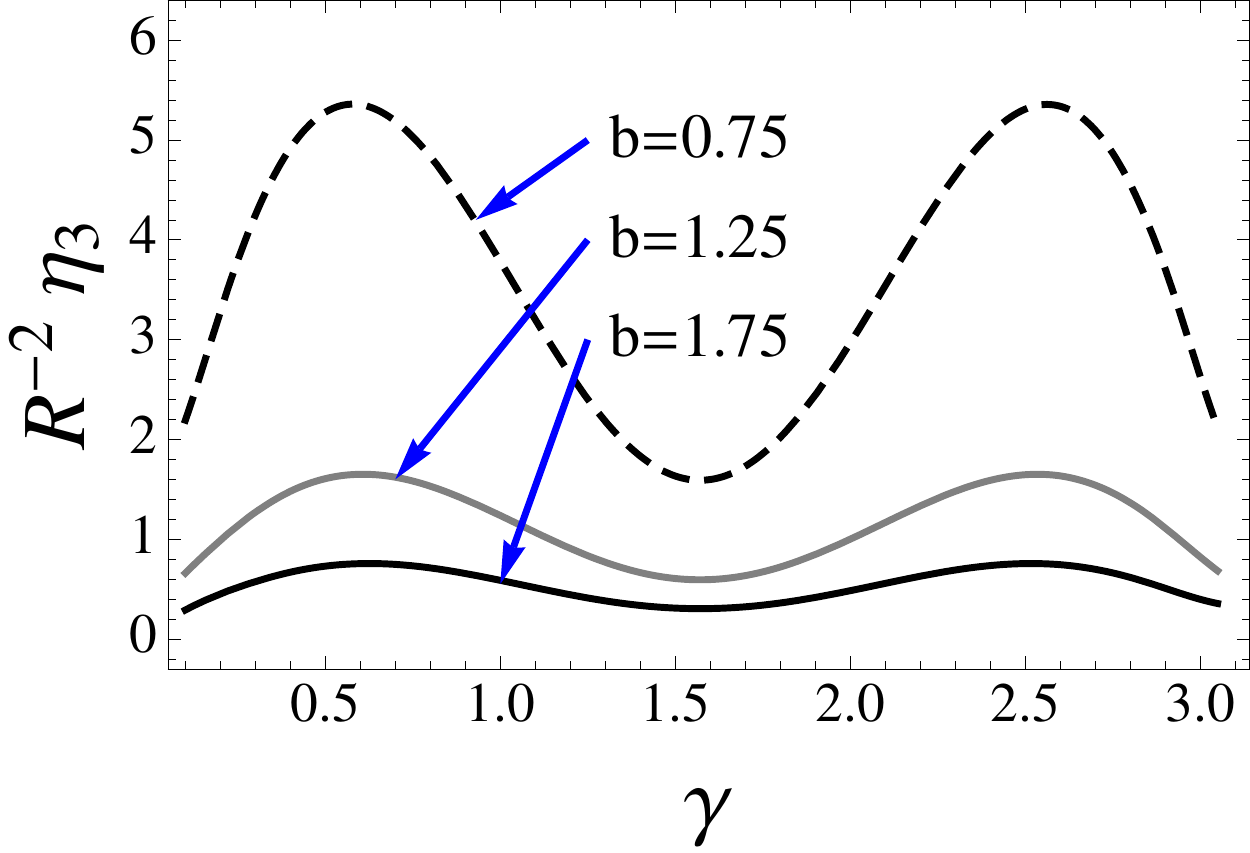}
\caption{Components of the dissipative tensor $\eta_1$ and $\eta_3$, right and left panel respectively, as a function of the helicity for different values of the adimensionalized external magnetic field $b$. The dashed, grey and black curves are obtained for $b=0.75$, $b=1.25$ and $b=1.75$, respectively.}
\label{fig:eta1_2}
\end{center}
\end{figure} 

The curves shown in Fig. \ref{fig:eta1_2} present the result obtained for the helicity-dependent parts of the non-zero components of the dissipative tensor, $\eta_1$ and $\eta_3$. The plots were performed for several values of the external magnetic field $b$. It is observed that as the magnetic field is decreased, the variation of the components of the dissipative tensor is enhanced. From Eqs. (\ref{eq:eta1}) and (\ref{eq:eta3}), it is clear that this is caused by the increase in size of the skyrmion profile due to its strong dependence on the external magnetic field (see Fig. \ref{fig:skyrmiontextures}({f}) and also Fig. \ref{fig:skyrmiontextures}({a-d})).

\subsection{Generalized Force and Equations of Motion}

In order to compute the generalized force we first need to evaluate the magnetic energy for the collective coordinate-dependent skyrmion texture $\bs{\Omega}(\bs{r} - \bs{r}_0, \gamma)$. The magnetic energy thus obtained, now a function of the collective coordinates, reads $\mathcal{U}(\bs{r}_0, \gamma) =  \mathcal{U}_{\text{DM}}(\bs{r}_0, \gamma) + \mathcal{U}_{\text{EX-Z}}(\gamma)$, where $\mathcal{U}_{\text{DM}}(\bs{r}_0, \gamma) = 2dM^2\sin\gamma\int d^2\bs{r} D(\bs{r}_0 + \bs{r}) f(r, \gamma)$, $\mathcal{U}_{\text{EX-Z}}(\gamma) = dJM^2g(\gamma)$, and
\begin{eqnarray}
f(r, \gamma) &\label{eq:f}=& \frac{\partial\Theta(r, \gamma)}{\partial r} + \frac{\sin2\Theta(r, \gamma)}{2r},\\
\nonumber g(\gamma) &\label{eq:g}=& 2\pi\int_0^\infty dr r \left[ \left( \frac{\partial\Theta(r, \gamma)}{\partial r} \right)^2 + \frac{\sin^2\Theta(r, \gamma)}{r^2} \right.\\
&& \left.\phantom{2\pi\int_0^\infty dr r} - \frac{\bar{B}}{J}\cos\Theta(r, \gamma) \right].
\end{eqnarray}
It is worth noting that while the exchange and Zeeman energy terms become helicity-dependent, $\mathcal{U}_{\text{EX-Z}}(\gamma)$, the DMI energy also acquires a dependence on the skyrmion position, $\mathcal{U}_{\text{DM}}(\bs{r}_0, \gamma)$. This was expected since only the DMI coupling was position-dependent. 

As outlined in section \ref{sec:Coordinates}, the components of the generalized force follow from straightforward partial differentiation of $\mathcal{U}(\bs{r}_0, \gamma)$. For the sake of clarity, hereafter we will use the collective coordinate symbols as subindices for the generalized force components instead of numbers. Thus, $F_{\{1,2\}} = - \partial\mathcal{U}/\partial \{ X, Y \} = F_{\{ X, Y \}}$, and $F_3 = - \partial\mathcal{U}/\partial \gamma = F_\gamma$. For a DMI coupling with arbitrary spatial dependence on the thin film plane, $D(\bs{r})$, the generalized force components are given below
\begin{eqnarray}
\label{eq:FXY}F_{\{ X, Y \}} &=& - 2dM^2\sin\gamma\int d^2\bs{r} \frac{\partial D(\bs{r}_0 + \bs{r})}{\partial \{ X, Y \}} f(r, \gamma),\\
\nonumber F_\gamma & = & - dM^2 \frac{\partial}{\partial\gamma} \left[ 2\sin\gamma\int d^2\bs{r} D(\bs{r}_0 + \bs{r})f(r, \gamma) \right.\\
\label{eq:Fgamma}&& \left.\phantom{- dM^2 \frac{\partial}{\partial\gamma}} + J g(\gamma) \right].
\end{eqnarray}

Using the explicit expressions determined in the previous section for the gyrotropic and dissipative tensors, we can write the Generalized Thiele's Equation for our model as the following autonomous, third-order dynamical system
\begin{eqnarray}
\label{eq:GTX}\dot{Y} - \alpha\eta_1(\gamma)\dot{X} + \frac{1}{4\pi d\ssf{J}}F_X(X,Y,\gamma) &=& 0,\\
\label{eq:GTY}\dot{X} + \alpha\eta_1(\gamma)\dot{Y} - \frac{1}{4\pi d\ssf{J}}F_Y(X,Y,\gamma) &=& 0,\\
\label{eq:GTg}\alpha\eta_3(\gamma)\dot{\gamma} - \frac{1}{4\pi d\ssf{J}}F_\gamma(X,Y,\gamma) &=& 0.
\end{eqnarray}
Although expected when internal degrees of freedom associated with the change of shape of the skyrmion are incorporated, our equations do not contain an inertia term. Unlike the results presented in Ref. [\onlinecite{Makhfudz}], where by integrating out the internal degrees of freedom an inertia term arises, the focus of our work is to explicitly track the helicity dynamics. If Eq. \eqref{eq:GTg} were used to eliminate $\gamma$ from Eqs. \eqref{eq:GTX} and \eqref{eq:GTY}, an inertia term, among others, should emerge.

Now that we have constructed the Generalized Thiele's Equation which describes the effective skyrmion dynamics in thin films with an arbitrary DMI coupling, we will focus exclusively on the special case of an engineered DMI coupling with one-dimensional spatial dependence. Besides being one of the simplest, we are interested in this particular case because skyrmion-supporting samples with this feature have already been reported. Without loss of generality, we choose the $x$-axis as the direction along which the DMI coupling varies, i.e., $D(\bs{r}) = D(x)$. This choice makes the magnetic energy, and consequently the generalized force, no longer dependent on $Y$. Our model can be further simplified by eliminating $\dot{Y}$ using \eqref{eq:GTY}---now with $F_Y = 0$---giving way to the following adimensionalized second-order dynamical system
\begin{align}
\frac{d\tilde{X}}{d\tau}&\label{eq:GenThielEq1} = \frac{\alpha\tilde{\eta}_1(\gamma)}{1+\alpha^2\tilde{\eta}^2_1(\gamma)}\tilde{F}_X(\tilde{X},\gamma),\\
\frac{d\gamma}{d\tau}&\label{eq:GenThielEq2} = \frac{1}{\alpha\tilde{\eta}_3(\gamma)}\tilde{F}_{\gamma}(\tilde{X},\gamma),
\end{align}
where all spatial variables and time have been scaled by $R = J/D$ and $T = \pi\ssf{J}R/(2DM^2)$, respectively. Here we have also introduced $\tilde{X}=X/R$, $\tilde{\eta}_1=\eta_1$, $\tilde{\eta}_3=\eta_3/R^2$, $\tilde{F}_X = F_X/(8dDM^2)$, and $\tilde{F}_{\gamma} = F_\gamma/(8dRDM^2)$. Therefore, only $\tilde{X}$ and $\gamma$ govern the effective skyrmion dynamics while $\tilde{Y} = Y/R$ is now a slave variable whose time evolution is determined from 
\begin{equation}\label{eq:GenThielEq3}
\frac{d\tilde{Y}}{d\tau} = - \tan\delta(\gamma)\frac{d\tilde{X}}{d\tau},
\end{equation}
where the Hall angle $\delta(\gamma)$ satisfies $\tan\delta(\gamma) = 1/[\alpha\tilde{\eta}_1(\gamma)]$. Because of the typical small values of the Gilbert damping, the Hall angle would attain an almost $\gamma$-independent value close to $\pi/2$, thus determining a large ratio between the $Y$- and $X$-velocities. However, if the variation of ${\tilde\eta}_1(\gamma)$ is made large enough, by decreasing the external magnetic field, it could be possible to observe the helicity dependence predicted for $\delta(\gamma)$.

The next two sections are devoted to the effective skyrmion dynamics in thin films, predicted by the model developed so far, in two particular scenarios of engineered DMI coupling. In both cases we will proceed to determine $\tilde{F}_X$ and $\tilde{F}_{\gamma}$, then the general Eqs. \eqref{eq:GenThielEq1}, \eqref{eq:GenThielEq2}, and \eqref{eq:GenThielEq3} will be used to analyze the resulting dynamics. 

\section{Half-Planes with opposite-sign DMI}

\begin{figure}[ht]
\begin{center}
\includegraphics[width=\columnwidth]{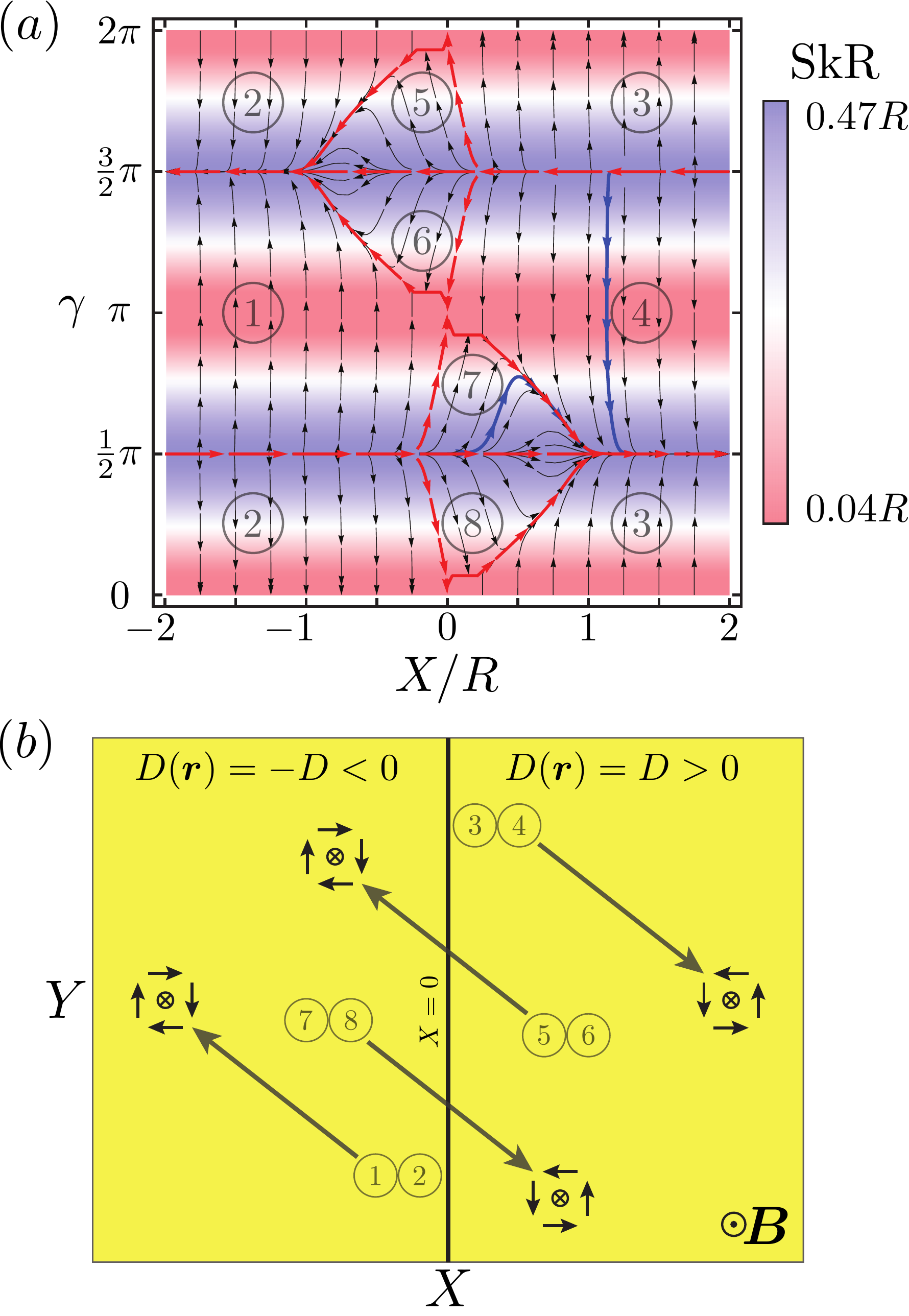}
\caption{(color online). (a) Phase portrait of solutions of the dynamical system for a skyrmion in a thin film split into two half-planes with opposite-sign DMI couplings, for $\alpha = 0.5$ and $b = 2$. The color code represents the skyrmion radius, SkR, defined as that for which $\theta = \frac{\pi}{2}$. Highlighted in red are the separatrices which determine the eight labeled regions. While trajectories from regions 1-4 always remain within either of the two domains, regions 5-8 host trajectories that can cross over. (b) Schematic of a the eight types of trajectories allowed by this dynamical system.}
\label{fig:OppDMI1}
\end{center}
\end{figure}

\begin{figure}[ht]
\begin{center}
\includegraphics[width=\columnwidth]{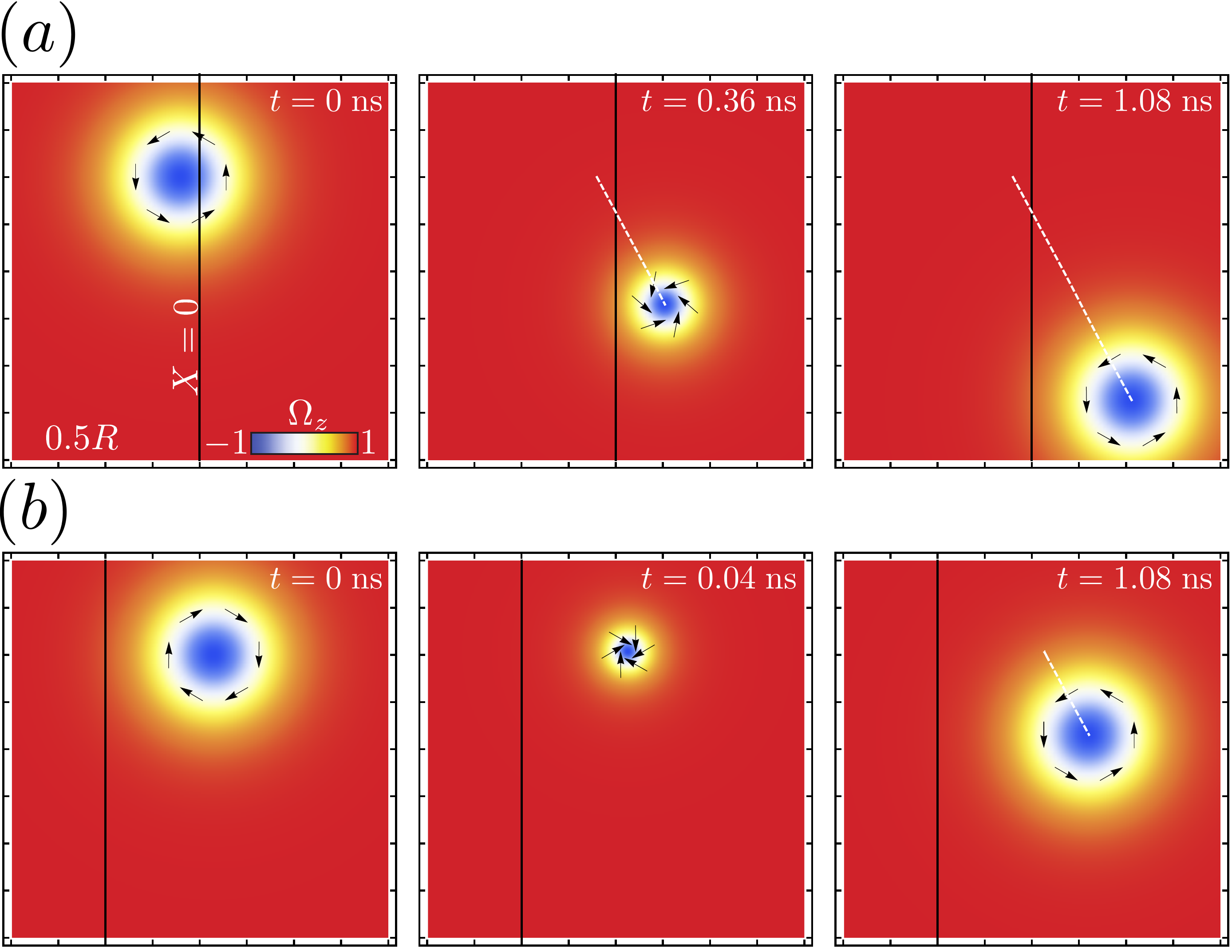}
\caption{(color online). Snapshots of two select trajectories, highlighted in blue, from the phase portrait in Fig. \ref{fig:OppDMI1}. (a) From region 7, a skyrmion crosses over the interface, at $X = 0$, between the two domains. (b) From region 4, a skyrmion reverts its helicity. Using parameter values for Fe$_{0.5}$Co$_{0.5}$Si, the time scale introduced to adimensionalize the dynamical system was estimated as $T = 0.36\;\mbox{ns}$.}
\label{fig:OppDMI2}
\end{center}
\end{figure}

Thin films of Mn$_{1-x}$Fe$_x$Ge with an engineered, spatially-varying DMI strength have been recently synthesized \cite{Shibata}. LTEM measurements revealed the existence of neighboring domains hosting skyrmions with opposite helicity: $\gamma = \frac{\pi}{2}$ and $\gamma = \frac{3\pi}{2}$, respectively. Skyrmion dynamics in such systems can be modeled by considering a thin film split into two half-planes with opposite-sign DMI couplings. If the interface between the two domains is located at $x=0$, the spatial dependence of the DMI coupling can be written as $D(\bs{r}) = D\;\mbox{sgn}(x)$. Using Eqs. \eqref{eq:FXY} and \eqref{eq:Fgamma}, the adimensionalized components of the generalized force are calculated to be
\begin{eqnarray}
\tilde{F}_X(\tilde{X},\gamma) &=& - \sin\gamma\tilde{f}_1(\tilde{X}, \gamma),\\
\tilde{F}_{\gamma}(\tilde{X},\gamma) &=& - \frac{\partial}{\partial\gamma}\left[ \sin\gamma\tilde{f}_3(\tilde{X}, \gamma) + \tfrac{1}{8}\tilde{g}(\gamma) \right],
\end{eqnarray}
where
\begin{eqnarray}
\tilde{f}_1(\tilde{X}, \gamma) &=& \int_0^\infty dv \tilde{f}(\sqrt{\tilde{X}^2 + v^2}, \gamma),\\
\tilde{f}_3(\tilde{X}, \gamma) &=& \int_0^\infty dv\int_0^{\tilde{X}} du \tilde{f}(\sqrt{u^2 + v^2}, \gamma),\\
\nonumber\tilde{g}(\gamma) &=& 2\pi\int_0^\infty d\rho\rho \left[ \left( \frac{\partial\theta(\rho,\gamma)}{\partial\rho} \right)^2 + \frac{\sin^2\theta(\rho,\gamma)}{\rho^2} \right.\\
&& \left. \phantom{\quad\quad\quad\quad\frac{1}{2}} - 2b\cos\theta(\rho,\gamma) \right].
\end{eqnarray}
The dimensionless functions introduced above, $\tilde{g}(\gamma) = g(\gamma)$ and $\tilde{f}(\rho,\gamma) = Rf(r,\gamma)$, were obtained by scaling the radial coordinate as $\rho=r/R$, and then defining $\theta(\rho,\gamma) = \Theta(R\rho,\gamma)$.

In samples with uniform DMI coupling $D$, the static, axially-symmetric skyrmions that extremize the magnetic free energy in Eq. \eqref{eq:MagneticEnergy} have helicity $\frac{\pi}{2}$ or $\frac{3\pi}{2}$ for $D > 0$ and $D < 0$, respectively. A skyrmion with either of those helicities does not extremize the magnetic energy in the sample with nonuniform DMI coupling considered in this section, unless it is located at an infinite distance away from the interface. Therefore, the expected long-term dynamics of a skyrmion in this type of sample is to flow toward $(\tilde{X} = \infty, \gamma = \pi/2)$ or $(\tilde{X} = -\infty, \gamma = 3\pi/2)$ depending on its initial condition. Indeed, as depicted in the phase portrait in Fig. \ref{fig:OppDMI1} (a), these are the only two attractors---stable fixed points---of the corresponding dynamical system. The unstable fixed points are $(\tilde{X} = 0, \gamma = 0, 2\pi)$, $(\tilde{X} = 0, \gamma = \pi)$. The separatrices, i.e., trajectories that separate regions of the phase portrait with qualitatively different solutions, are also shown, in red. A total of eight qualitatively different possible solutions emerge from the structure of fixed points and separatrices of this dynamical system. These correspond to the eight regions labeled in the phase portrait. A diagram sketching all types of possible trajectories in the $XY$ plane is included in Fig. \ref{fig:OppDMI1} (b). 

According to the phase portrait of solutions, a skyrmion that has been nucleated to the right of the interface ($\tilde{X} > 0$) could either remain within this domain and eventually flow away from the interface toward the attractor $(\tilde{X} = \infty, \gamma = \pi/2)$ or cross over to the left domain and flow asymptotically to $(\tilde{X} = -\infty, \gamma = 3\pi/2)$. Among the possible trajectories that remain within the right domain, those whose initial condition is close to the interface  could have the skyrmion approach the interface first and the turn around to finally flow toward the attractor. Interestingly, the time evolution of the helicity is strictly monotonic (clockwise/counterclockwise as in region 3/4 [Fig. \ref{fig:OppDMI1} (a)]) for those skyrmions that remain to the right of the interface, whereas the helicity of skyrmions that cross over to the left domain can increase/decrease to a maximum/minimum value before decreasing/increasing to $\frac{3\pi}{2}$ at the attractor. A similar analysis can be made for skyrmions nucleated to the left of the interface ($\tilde{X} < 0$).

Two features observed for skyrmions nucleated at large distances from the interface are worth noting. First, the further a skyrmion is nucleated from the interface, the smaller becomes its probability to cross over. This can be understood by observing how little regions 5-8 [Fig. \ref{fig:OppDMI1} (a)], which host solutions that cross over, such as that shown in Fig. \ref{fig:OppDMI2} (a), extend away from the interface. Second, for those initial conditions that do not result in the skyrmion crossing over [Fig. \ref{fig:OppDMI2} (b)], the time evolution  appears to take place in two steps: a fast change in the helicity toward its asymptotic value with almost no change in the skyrmion position, followed by a slow flow of the skyrmion away from the interface.   

\section{Linearly varying DMI}

The skyrmion dynamics under a linear gradient of the DMI coupling is considered in this section. The spatially-dependent DMI coupling consists of a linear variation along the $x$-coordinate as $D(\bs{r}) = D+D_0x/l$ for $0<x<l$, connecting two regions at $x<0$ and $x>l$ where $D(\bs{r}) = D$ and $D(\bs{r}) = D+D_0$, respectively. The coupling $D$ is assumed positive in order to fix the skyrmion profile with helicity $\gamma=\pi/2$ far from the transition region of $D(\bs{r})$. As we previously mentioned in Sec. II, the spatial change in the DMI behaves as external forces acting on the helicity and center of mass position of the magnetic skyrmion. Thus, the generalized forces along the coordinates $\tilde{X}$ and $\gamma$ explicitly read, 
\begin{align}
\tilde{F}_X[\tilde{X},\gamma]&\label{eq:GForce_X}=-\sin\gamma\frac{D_0}{4D} \tilde{\mathsf{I}}_1[\tilde{X},\gamma],\\
\tilde{F}_{\gamma}[\tilde{X},\gamma]&\label{eq:GForce_gamma}\nonumber=-\frac{1}{4}\frac{\partial}{\partial \gamma}\left[\tilde{\cal E}^0_{DM}(\gamma)+\sin\gamma \frac{D_0}{D}\left(\tilde{X}\tilde{\mathsf{I}}_1[\tilde{X},\gamma]\right.\right.\\
&\left.\left.\qquad+\tilde{\mathsf{I}}_2[\tilde{X},\gamma]+\tilde{\mathsf{I}}_3[\tilde{X},\gamma]\right)+2\pi{g}(\gamma)\right],
\end{align}
where the dimensionless integrals $\tilde{\mathsf{I}}_1$, $\tilde{\mathsf{I}}_2$, $\tilde{\mathsf{I}}_3$ and $\tilde{\cal E}^0_{DM}$ entering in the above expressions are defined as follows 
\begin{align}
\tilde{\mathsf{I}}_1[\tilde{X},\gamma]&\label{eq:IntI1}=\frac{2}{\lambda}\int^{\infty}_{0}d\tilde{y}'\int^{\lambda-\tilde{X}}_{-\tilde{X}} d\tilde{x}'\tilde{f}({\rho}',\gamma),\\
\tilde{\mathsf{I}}_2[\tilde{X},\gamma]&\label{eq:IntI2}=\frac{2}{\lambda}\int^{\infty}_{0}d\tilde{y}'\int^{\lambda-\tilde{X}}_{-\tilde{X}} d\tilde{x}' \tilde{x}'\tilde{f}(\rho',\gamma),\\
\tilde{\mathsf{I}}_3[\tilde{X},\gamma]&\label{eq:IntI3}=2\int^{\infty}_{0}d\tilde{y}'\int^{\infty}_{\lambda-\tilde{X}}d\tilde{x}' \tilde{f}(\rho',\gamma),\\
\tilde{{\cal E}}^0_{DM}(\gamma)&\label{eq:IntI4}=2\pi \sin\gamma\int^{\infty}_{0} d\rho' \rho' \tilde{f}(\rho',\gamma).
\end{align}
where $\tilde{f}(\rho,\gamma)$ was already introduced in Sec. III and $\lambda=l/R$ is the ratio between the scale of variation of the DMI coupling and the relevant magnetic length scale. 

Previous to explore the case devoted to this section, we will analyze first the uniform DMI coupling case, i.e., in the limit when the DMI gradient $D_0$ goes to zero. This assumption implies that the force $\tilde{F}_X=0$ and thus no motion of the center of mass of the skyrmion is expected. On the other hand, the force acting on the helicity turns out to derive from the potential energy ${\cal U}[D_0=0]=\left(2\pi{g}+\tilde{\cal E}^0_{DM}\right)/4$, which is symmetric with a local minimum at $\gamma=\pi/2$. Therefore, the time evolution of the skyrmion helicity is driven to the equilibrium configuration at $\gamma=\pi/2$, for any given initial condition. The same argument holds for a negative DMI coupling, having the equilibrium skyrmion profile a helicity $\gamma=3\pi/2$. A similar behavior is observed for the case treated in Sec. III, which occurs far from transition region located at $\tilde{X}=0$.

\begin{figure}[ht]
\begin{center}
\includegraphics[width=3.5in]{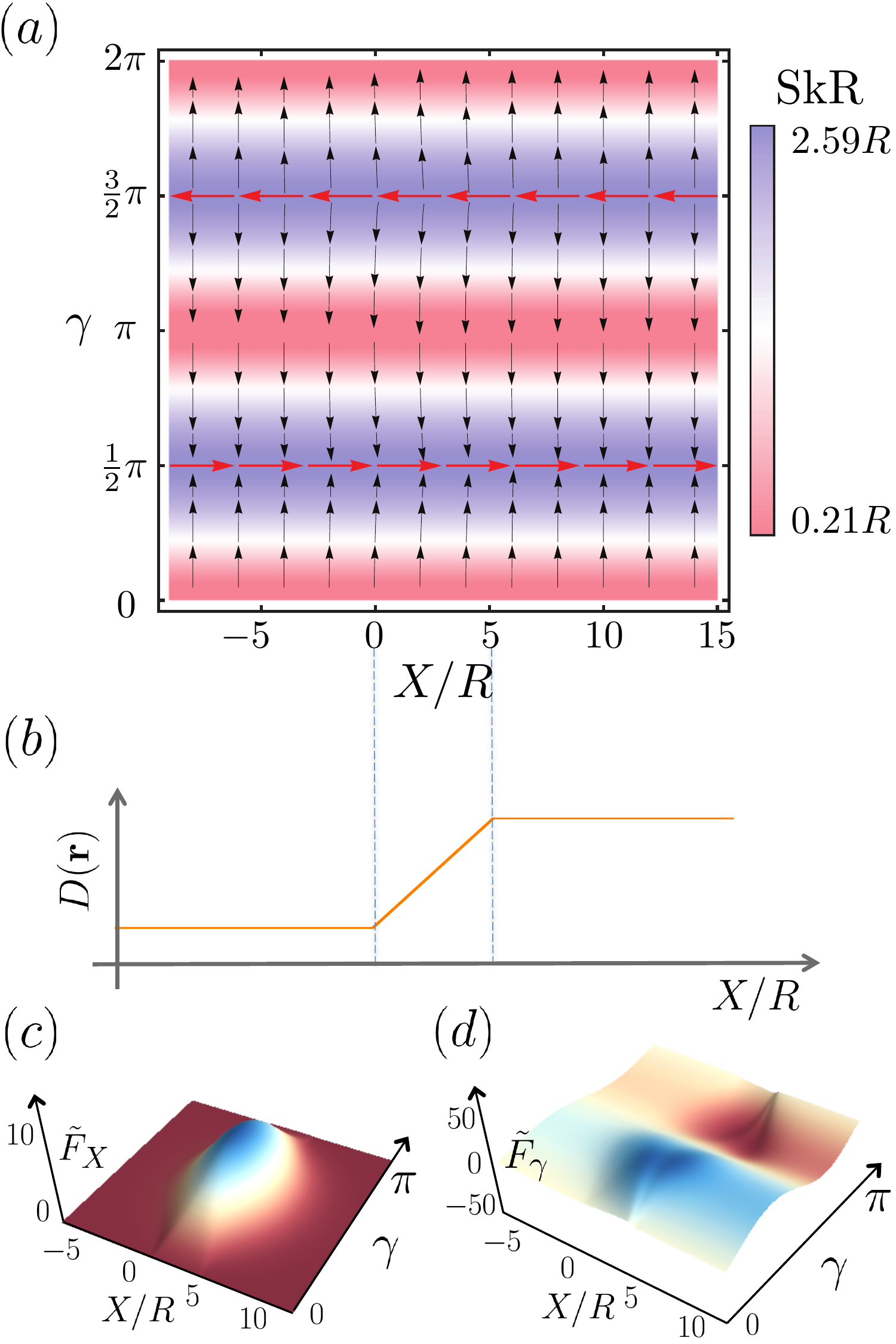}
\caption{(color online). (a) Phase portrait representation for skyrmion dynamics under the linear gradient of DMI coupling (panel (b)). The color code represents the skyrmion radius, SkR, defined as that for which $\theta = \frac{\pi}{2}$. The flow of trajectories is plotted as a function of the center of mass position $X/R$ and helicity $\gamma$ for an external magnetic field $b=0.5$, and for the parameters given by $D_0/D=5$, $\lambda=5$ and $\alpha=0.5$. Red lines, located at $\gamma=\pi/2$ and $\gamma=3\pi/2$, refer to those stable and unstable solutions in the phase portrait, respectively. In (c) and (d), the behavior of generalized forces $\tilde{F}_X$ and $\tilde{F}_{\gamma}$ are displayed. Taking advantage of their symmetric behavior, the plots were done in the range for the helicity between $0$ and $\pi$.}
\label{fig:PhasePortraitLinearDMI}
\end{center}
\end{figure} 

In Fig. \ref{fig:PhasePortraitLinearDMI}{(a)} the phase portrait representation for the skyrmion dynamics under the linear gradient of DMI coupling is shown, (see Fig. \ref{fig:PhasePortraitLinearDMI}{(b)}). The flow of trajectories obeying the generalized Thiele's equation are displayed as a function of the helicity $\gamma$ and the center of mass position $\tilde{X}$. The computation was carried out by solving the set of integrals Eqs. (\ref{eq:IntI1}-\ref{eq:IntI4}), for the parameter $\lambda=5$ and a skyrmion profile solution obtained for $b=0.5$. Alongside this, the generalized forces were determined for the parameters $D_0/D=5$ and $\alpha=0.5$, whose behavior is displayed in Fig. \ref{fig:PhasePortraitLinearDMI}{(c)} and {(d)}. As was explained in Sec. III, at every single point on each trajectory in Fig. \ref{fig:PhasePortraitLinearDMI}{(a)} there is a vector whose components represent the value for $\left({d\tilde{X}}/{d\tau},{d\gamma}/{d\tau}\right)$. Indeed, for a set of initial conditions dictated by $\gamma\neq 3\pi/2$ we see that all the trajectories converge to the solution for which the helicity is $\gamma=\pi/2$, represented by the right-directed red line. The latest, is because $D$, as well as $D_0$, have been assumed positive and therefore, the  skyrmion configuration for which the magnetic energy is extremized is that with helicity $\pi/2$. In particular, the motion for a helicity $\gamma=\pi/2$ (bottom red line) occurs in the rightward direction, i.e. {opposite} to the gradient of the DMI coupling. The above statement ($\tilde{F}_X>0$) comes from the fact that the DMI energy ${\cal U}_{\text{DM}}$ is strictly negative, which is a consequence of $f(r,\gamma)<0$, see Eq. (\ref{eq:f}). Furthermore, the evolution at $\gamma=\pi/2$ the skyrmion towards the rightward region preserves its size, which is the same as the trajectory at $\gamma=3\pi/2$. Indeed, this is because the computation of the skyrmion profile (for more details see Appendix \ref{appendix}) is accomplished as a function of the helicity and external magnetic field, but does not take into account the local variation of the DMI. Therefore, for those trajectories where the helicity is constant, there will be no change in the size or shape of the skyrmion. 

Finally, it is worth commenting about the connection between the linear case treated here and that described in Sec. III. The half-planes with opposite-sign DMI case can be obtained by taking an appropriate limit to the generalized forces $\tilde{F}_X$ and $\tilde{F}_{\gamma}$ described in this section. In fact, assuming that $D<0$, choosing that $D_0=-2\left|D\right|$ and taking the limit when $\lambda\rightarrow 0$, we make sure that the linear transition occurs between opposite values in a very narrow region. Under these considerations, the integral expressions for $\mathsf{I}_1$ and $\mathsf{I}_2$ turn out to obey $\tilde{\mathsf{I}}_1[\tilde{X},\gamma]=2\tilde{f}_1(\tilde{X}, \gamma) $ and $\tilde{\mathsf{I}}_2[\tilde{X},\gamma]=-2\tilde{X}\tilde{f}_1(\tilde{X}, \gamma)$. Therefore, the results for $\tilde{F}_X$ and $\tilde{F}_{\gamma}$ given by Eqs. (\ref{eq:GForce_X}) and (\ref{eq:GForce_gamma}) reduce to those defined in Sec. III.

\section{Discussion and Conclusions}

We have investigated single skyrmion dynamics in chiral magnets with locally varying Dzyaloshinskii-Moriya interactions. An effective description, based on the collective coordinates method, was established in terms of the skyrmion center of mass and helicity, which were found to satisfy a generalized Thiele's equation. In the construction of this equation, only the static skyrmion radial profile, $\Theta(r,\gamma)$ was needed to compute the gyrotropic and the dissipative tensors. The spatially dependent DMI coupling only entered in the calculation of the generalized forces. This separation of dependencies was exploited to write a general dynamical system governing the skyrmion dynamics that could be used for the two cases discussed in the previous sections. Therefore, the dynamical system to be solved in each case was fully determined once the DMI coupling was specified so the generalized forces could be calculated. It should be noted that although the radius of the skyrmion was not included as a collective coordinate, by using a helicity-dependent skyrmion radial profile to derive the generalized Thiele's equation, the size of the skyrmion inherited its dynamics from the helicity. This is a general feature of our model independent of the particular DMI coupling considered, hence it was observed in the two studied cases. Another feature shared by both cases is the helicity dependence of the Hall angle. Since $\delta(\gamma)$ derives from the dissipative tensor, it will be the same for any form of the DMI coupling. As with the skyrmion radius, the Hall angle also inherits its time dependence from the helicity. Furthermore, for sufficiently small values of the Gilbert damping, the Hall angle becomes almost independent of the helicity with a value close to $\pi/2$. 

As evidenced by their respective phase portraits, the dynamics of the case from section III is richer than that from section IV. Due to the choice of parameters in the case of linearly varying DMI, only one attractor was present and two types of qualitatively different trajectories, differing by the sense of rotation of the helicity, were allowed. On the other hand, by virtue of the change of sign in the DMI coupling, the other case had two attractors as well as eight qualitatively different possible trajectories.

The simple helicity dynamics allowed by the two cases we have investigated, should not be seen as mere academic exercises. On the contrary, the proposed engineered DMI coupling scenarios must be considered as the simplest testbeds where the nontrivial time evolution of the skyrmion helicity can be studied in a controlled fashion. Indeed, the features observed in the two cases discussed could be detected by the experimental technique used in Ref. [\onlinecite{Onose}] where the low-energy dynamics of a skyrmion lattice was investigated measuring its microwave response.

In this work, we predicted how a single magnetic skyrmion texture adjusts its helicity as it propagates under a spatially modulated DMI. The results presented here establish the road to single magnetic skyrmion manipulation with full control over all its relevant degrees of freedom. Moreover, we expect future applications to take advantage of the relation between the skyrmion position and helicity using the structure of their phase portrait as a valuable guide. Finally, our findings constitute the first step toward a more complete understanding of the physics of the skyrmion helicity, which opens numerous opportunities for an effective control of skyrmions in future high-density storage and logic devices. Nucleation of skyrmions in chiral magnets with inhomogeneous DMI as well as the role of electric currents on their dynamics, will be addressed elsewhere.

\begin{acknowledgments}
It is a pleasure to thank A. S. Nu\~nez and R. A. Duine by fruitful and stimulating discussions. R. E. T. acknowledges support by Proyecto Fondecyt Postdoctorado number 3150372. S.D. acknowledges partial support from the International Fulbright Science and Technology Award.
\end{acknowledgments}

\appendix

\section{Skyrmion Radial Profile}\label{appendix}

The radial profile of a static, axially symmetric skyrmion with helicity $\gamma$, $\Theta(r, \gamma)$, was at the core of our calculations for the terms in the Generalized Thiele's Equation. Assuming the skyrmion was nucleated in a region with a locally uniform DMI coupling, the magnetic energy Eq. \eqref{eq:MagneticEnergy}, with $D(\bs{r}) = D$, was evaluated in $\bs{M} = M\bs{\Omega}(\bs{r}, \gamma)$ with $\bs{\Omega}(\bs{r}, \gamma)$ as in Eq. \eqref{eq:StaticSk}. Thus, a functional in the skyrmion radial profile was obtained. Extremizing this functional led to the following Euler-Lagrange equation 
\begin{align}\label{eq:skyrmionprofile}
\frac{d^2\theta}{d\rho^2} + \frac{1}{\rho}\frac{d\theta}{d\rho} + \sin\gamma\frac{2\sin^2\theta}{\rho} - \frac{\sin 2\theta}{2\rho^2} - b\sin\theta=0,
\end{align}
with boundary conditions $\theta(0,\gamma)=\pi$ and $\theta(\infty,\gamma)=0$. The above ODE has been adimensionalized by scaling the radial coordinate as $\rho=r/R$, where $R=J/D$, as well as defining $\theta(\rho,\gamma) = \Theta(R\rho,\gamma)$ and $b = \frac{\bar{B}R^2}{2J}$. It is worth pointing out that the helicity enters in Eq. \eqref{eq:skyrmionprofile} as a parameter, in the same way as the adimensionalized external magnetic field $b$. Only after solving this equation numerically, using the shooting method, $\gamma$ was promoted to collective coordinate status.

The skyrmion radius is defined as that for which $\theta = \frac{\pi}{2}$. To account for the underlying lattice structure, we do not allow skyrmions with a radius smaller than $\rho^* = \sqrt{2}a/R$, where $a$ is the lattice constant. This restriction is enforced demanding that for $\gamma \in [0,\gamma^*]\cup[\pi - \gamma^*,\pi + \gamma^*]\cup[2\pi - \gamma^*,2\pi]$, $\theta(\rho,\gamma) = \theta(\rho,\gamma^*)$, where $\gamma^*$ is such that $\theta(\rho^*,\gamma^*) = \frac{\pi}{2}$.

Schematics of the skyrmion magnetization field obtained are depicted in Fig. \ref{fig:skyrmiontextures} for several values of the helicity (panels {(a)-(d)}). The dependence of the skyrmion radial profile on the external magnetic field and the helicity are shown in Fig. \ref{fig:skyrmiontextures}{(e)} and Fig. \ref{fig:skyrmiontextures}{(f)}, respectively.

\bibliographystyle{elsarticle-num}

\begin{thebibliography}{100}
\bibitem{Skyrme}T. H. R. Skyrme, Nucl. Phys. {\bf 31}, 556 (1962).
\bibitem{Bogdanov}A.N. Bogdanov and D.A. Yablonskii, Zh. Eksp. Teor. Fiz. {\bf 95},
178 (1989) [Sov. Phys. JETP {\bf 68}, 101 (1989)]; A. Bogdanov and A. Hubert, J. Magn. Magn. Mater. {\bf 138}, 255 (1994); A. N. Bogdanov, U. K. Rossler and A. A. Shestakov  Phys. Rev. E {\bf 67}, 016602 (2003); U. K. Rossler, N. Bogdanov, and C. Pfleiderer, Nature {\bf 442}, 797 (2006).
\bibitem{nagaosa}N. Nagaosa and Y. Tokura, Nature Nanotechnology {\bf 8}, 899 (2013).
\bibitem{Fert1}A. Fert,	V. Cros and J. Sampaio, Nature Nanotechnology {\bf 8}, 152 (2013); R. Duine, Nature Nanotechnology {\bf 8}, 800 (2013).


\bibitem{Muhlbauer}S. Muhlbauer, B. Binz, F. Jonietz, C. Pfleiderer, A. Rosch, A. Neubauer, R. Georgii, P. Boni, Science {\bf 323}, 915 (2009).
\bibitem{Jonietz} F. Jonietz , S. Muhlbauer, C. Pfleiderer, A. Neubauer, W. Munzer, A. Bauer, T. Adams, R. Georgii, P. Boni, R. A. Duine, K. Everschor, M. Garst, A. Rosch, Science {\bf 330}, 1648 (2010).
\bibitem{Neubauer}A. Neubauer, C. Pfleiderer, B. Binz, A. Rosch, R. Ritz, P. G. Niklowitz and P. Boni, Phys. Rev. Lett. {\bf 102}, 186602 (2009).

\bibitem{Munzer} W. Munzer, A. Neubauer, T. Adams, S. Muhlbauer, C. Franz, F. Jonietz, R. Georgii, P. Boni, B. Pedersen, M. Schmidt, A. Rosch and C. Pfleiderer, Phys. Rev. B {\bf 81}, 041203(R) (2010).
\bibitem{PMilde}P. Milde, D. Kuhler, J. Seidel, L. M. Eng, A. Bauer, A. Chacon, J. Kindervater, S. Muhlbauer,
 C. Pfleiderer, S. Buhrandt, C. Schutte and A. Rosch, Science {\bf 340}, 1076 (2013).
 \bibitem{Pfleiderer}C. Pfleiderer, T. Adams, A. Bauer, W. Biberacher, B. Binz, F. Birkelbach, P. Boni, C. Franz, R. Georgii, M. Janoschek, F. Jonietz, T. Keller, R. Ritz, S. Muhlbauer, W. Munzer, A. Neubauer, B. Pedersen and A. Rosch, J. Phys. Condens. Matter {\bf 22}, 164207 (2010).
 \bibitem{Yu2} X. Z. Yu, Y. Onose, N. Kanazawa, J. H. Park, J. H. Han, Y. Matsui, N. Nagaosa and Y. Tokura, Nature Materials {\bf 465}, 901 (2010).
\bibitem{Shibata}K. Shibata, X. Z. Yu, T. Hara, D. Morikawa, N. Kanazawa, K. Kimoto, S. Ishiwata, Y. Matsui and Y. Tokura, Nature Nanotechnology {\bf 8}, 723 (2013).

\bibitem{Yu1} X. Z. Yu, N. Kanazawa, Y. Onose, K. Kimoto, W. Z. Zhang, S. Ishiwata, Y. Matsui	 and Y. Tokura, Nature Materials {\bf 10}, 106 (2011).
\bibitem{Nagao} M. Nagao, Y. So, H. Yoshida, M. Isobe, T. Hara, K. Ishizuka and K. Kimoto, Nature Nanotechnology {\bf 8}, 325 (2013).

\bibitem{Seki}S. Seki, X. Z. Yu, S. Ishiwata and Y. Tokura, Science {\bf 336}, 198 (2012).

\bibitem{Heinze}S. Heinze, K. von Bergmann, M. Menzel, J. Brede, A. Kubetzka, R. Wiesendanger, G. Bihlmayer and S. Blugel, Nature Physics {\bf 7}, 713 (2011).


\bibitem{Shiomi}Y. Shiomi, N. Kanazawa, K. Shibata, Y. Onose, and Y. Tokura, Phys. Rev. B {\bf 88}, 064409 (2013).
\bibitem{Everschor}Karin Everschor, Markus Garst, R. A. Duine, and Achim Rosch, Phys. Rev. B {\bf 84}, 064401 (2011).
\bibitem{Berger}L. Berger, J. Appl. Phys. {\bf 55}, 1954 (1984); J. C. Slonczewski, J. Magn. Magn. Mater. {\bf 159}, L1 (1996).
\bibitem{Mochizuki}M. Mochizuki, X. Z. Yu, S. Seki, N. Kanazawa, W. Koshibae,	J. Zang, M. Mostovoy, Y. Tokura and N. Nagaosa, Nature Materials {\bf 13}, 241 (2014).
\bibitem{Fert} A. Fert, V. Cros and J. Sampaio, Nature Nanotechnology {\bf 8}, 152 (2013).
\bibitem{Troncoso}Roberto E. Troncoso and Alvaro S. N\'u\~nez, Phys. Rev. B {\bf 89}, 224403 (2014).
\bibitem{Reichhardt}C. Reichhardt, D. Ray, and C. J. Olson Reichhardt, Phys. Rev. B {\bf 91}, 104426 (2015); C Reichhardt, D Ray and C J Olson Reichhardt,  New Journal of Physics, {\bf 17} 073034 (2015).
\bibitem{Lin1}Shi-Zeng Lin, Cristian D. Batista, Charles Reichhardt, and Avadh Saxena, Phys. Rev. Lett. {\bf 112}, 187203 (2014); Lingyao Kong and Jiadong Zang, Phys. Rev. Lett. {\bf 111}, 067203 (2013).
\bibitem{Thiele} A. A. Thiele, Phys. Rev. Lett. {\bf 30}, 230 (1972).


\bibitem{DMI} I. Dzyaloshinskii, J. Phys. Chem. Solids {\bf 4}, 241 (1958); T. Moriya, Phys. Rev. {\bf 120}, 91 (1960).
\bibitem{Koshibae}Wataru Koshibae and Naoto Nagaosa, Nature Communications {\bf 5}, 5148 (2014).
\bibitem{Tsurkan} I. K\'ezsm\'arki, S. Bord\'acs, P. Milde, E. Neuber, L. M. Eng, J. S. White, H. M. R onnow, C. D. Dewhurst, M. Mochizuki, K. Yanai, H. Nakamura, D. Ehlers, V. Tsurkan, A. Loidl, Nature Materials (2015)  doi:10.1038/nmat4402.



\bibitem{Grigoriev} S. V. Grigoriev, D. Chernyshov, V. A. Dyadkin, V. Dmitriev, E. V. Moskvin, D. Lamago, Th. Wolf, D. Menzel, J. Schoenes, S. V. Maleyev, and H. Eckerlebe, Phys. Rev. B {\bf 81}, 012408 (2010); S. V. Grigoriev, N. M. Potapova, S. A. Siegfried, V. A. Dyadkin, E. V. Moskvin, V. Dmitriev, D. Menzel, C. D. Dewhurst, D. Chernyshov, R. A. Sadykov, L. N. Fomicheva, and A. V. Tsvyashchenko, Phys. Rev. Lett. {\bf 110}, 207201 (2013)

\bibitem{Grigoriev2}S. V. Grigoriev, D. Chernyshov, V. A. Dyadkin, V. Dmitriev, S. V. Maleyev, E. V. Moskvin,D. Menzel, J. Schoenes, and H. Eckerlebe, Phys. Rev. Lett. {\bf 102}, 037204 (2009).
\bibitem{Siegfried}S.A. Siegfried, E. V. Altynbaev, N. M. Chubova, V. Dyadkin, D. Chernyshov, E. V. Moskvin, D. Menzel, A. Heinemann, A. Schreyer, and S. V. Grigoriev, Phys. Rev. B {\bf 91}, 184406 (2015)
\bibitem{Morikawa} D. Morikawa, K. Shibata, N. Kanazawa, X. Z. Yu, and Y. Tokura, Phys. Rev. B {\bf 88}, 024408 (2013)

\bibitem{Koretsune}T. Koretsune, N. Nagaosa, R. Arita, Scientific Reports {\bf 5} 13302 (2015). 
\bibitem{Gayles}J. Gayles, F. Freimuth, T. Schena, G. Lani, P. Mavropoulos, R. Duine, S. Blugel, J. Sinova, Y. Mokrousov, Phys. Rev. Lett. {\bf 115}, 036602 (2015).
\bibitem{Chen}J. P. Chen, Y. L. Xie, Z. B. Yan, and J.M. Liu, J. Appl. Phys. {\bf 117}, 17C750 (2015).



\bibitem{JRoldan}A. Rold\'an-Molina, M. J. Santander, \'A. S. N\'u\~nez, J. Fern\'andez-Rossier, arXiv:1502.01950 [cond-mat.mes-hall].


\bibitem{Rajaraman}R. Rajaraman, Solitons and Instantons, North-Holland, 1982.
\bibitem{Tretiakov1}O. A. Tretiakov, D. Clarke, Gia-Wei Chern, Ya. B. Bazaliy, and O. Tchernyshyov, Phys. Rev. Lett. {\bf 100}, 127204 (2008).


\bibitem{Hals} Kjetil M. D. Hals and Arne Brataas, Phys. Rev. B {\bf 89}, 064426 (2014); M. E. Knoester, Jairo Sinova, and R. A. Duine, Phys. Rev. B {\bf 89}, 064425 (2014).

\bibitem{work in preparation}Sebasti\'an A. D\'iaz and Roberto E. Troncoso (to be published).


\bibitem{Makhfudz}Imam Makhfudz, Benjamin Kruger, and Oleg Tchernyshyov, Phys. Rev. Lett. {\bf 109}, 217201 (2012).


\bibitem{Onose}Y. Onose, Y. Okamura, S. Seki, S. Ishiwata, and Y. Tokura, Phys. Rev. Lett. {\bf 109}, 037603 (2012).


\end{thebibliography}

\end{document}